\documentclass[%
reprint,
superscriptaddress,
nofootinbib,
amsmath,amssymb,
aps,
prd,
floatfix,
superscriptaddress,nofootinbib,showpacs,longbibliography
]{revtex4-2}

\usepackage[utf8]{inputenc}
\usepackage{
	amsmath, 
	amssymb, 
	mathtools,
	xcolor,
	graphicx,
	listings,
	color, 
	dcolumn,
	multirow,
	hyperref,
	float,
	tabularx,
	soul,
	ulem
}

\hypersetup{
	colorlinks=true,
	linkcolor=blue,
	filecolor=magenta,      
	urlcolor=blue,
	citecolor=violet
}
\usepackage[toc,page]{appendix}
\usepackage[nolist]{acronym}

\definecolor{dkgreen}{rgb}{0,0.6,0}
\definecolor{gray}{rgb}{0.5,0.5,0.5}
\definecolor{mauve}{rgb}{0.58,0,0.82}
\sethlcolor{pink}
\newcolumntype{Y}{>{\centering\arraybackslash}X}
\begin{document}
	\title{Prospect of Constraining the EoS of Neutron Stars Using Post-Merger Signals}
	
	\author{Soham Mitra}\email{mitra.soham.2000@gmail.com}
	\affiliation{Indian Institute of Technology Bombay, Mumbai, Maharashtra, 400076, India}
	
	\author{Praveer Tiwari}\email{praveer.tiwari@iitb.ac.in}
	\affiliation{Indian Institute of Technology Bombay, Mumbai, Maharashtra, 400076, India}
	
	\author{Archana Pai}\email{archanap@iitb.ac.in}
	\affiliation{Indian Institute of Technology Bombay, Mumbai, Maharashtra, 400076, India}

	\date{\today}
		
	\begin{abstract}
		Post-merger gravitational-wave emission from a binary neutron star merger carries crucial information about the equation of state (EoS) of matter at high temperatures. Current gravitational wave detectors have limited sensitivities at post-merger frequencies in the range [1.5, 4] kHz. Therefore, valuable inferences can only be made after combining information from multiple \ac{BNS} events.  \citet{Criswell} carries out an injection study to infer the radius posterior for a $1.6 M_{\odot}$ NS by combining the information from injected BNS events via the hierarchical Bayesian inference (HBI) formulation. This formulation utilizes empirical relations that connect the peak frequency of the post-merger remnant with the chirp mass of the system, and the EoS proxy parameter $R_{1.6}$.  In this work, we extend the HBI formulation to other EoS proxy parameters (i.e., the radius of \ac{NS}, $R_{1.X}$,  with masses 1.2, 1.4, and 1.8 $M_{\odot}$) and combine the four $R_{1.X}$ posteriors through the piecewise polytropic EoS model to obtain measurable constraints on the EoS of the NS.  We show that the NS radii can be constrained to within $\sim$1 km ( $\sim$0.55 km) assuming uniform (astrophysical) prior on $R_{1.X}$ for injections in the A+ noise.  We also study systematic biases in the analysis coming from the limitations of empirical relations.   
	\end{abstract}
	
	\maketitle
	
	%  Define your achronyms here
	
\acrodef{BNS}[BNS]{binary neutron star}	
\acrodef{NS}[NS]{neutron star}
\acrodefplural{NS}[NSs]{neutron stars}
\acrodef{HBI}[HBI]{hierarchical Bayesian inference}
\acrodef{EM}[EM]{electromagnetic}
\acrodef{SPH}[SPH]{smooth particle hydrodynamics}
\acrodef{EoS}[EoS]{equation of state}
\acrodef{M-R}[M-R]{mass-radius}
\acrodef{M-lam}[M-$\Lambda$]{mass-tidal deformability}	
	
	\section{Introduction}
	\label{sec:Introduction}
	
	The detection of gravitational waves (GWs) from coalescing \ac{BNS} systems, GW170817 \citep{GW170817_Obs} and GW190425 \citep{LIGOScientific:2020aai}, by the LIGO-VIRGO collaboration, has put constraints on the composition of a cold non-rotating \ac{NS} via the \ac{EoS}.  { A number of studies~\cite{Huth_2022, dietrich2020multimessenger, pang2021nuclear, pang2022nmma, biswas2021towards, ghosh2022multi, Raaijmakers:2020, kat2020, biswas2022bayesian, pang2022nmma, pang2021nuclear, landry2020nonparametric, raaijmakers2021constraints, legred2021impact,biswas2021impact} attempted to characterize the NS interiors by combining these GW detections with other astrophysical ~\cite{Miller_1, Miller_2, Riley_1, Riley_2}, and terrestrial probes~\cite{russotto2016results, le2016constraining, hebeler2013equation, drischler2020well}. The information from BNS inspiral, however, can only probe the NS up until densities close to nuclear saturation \citep{LIGOScientific:2018hze}.} To probe deeper inside a \ac{NS}, we rely on the post-merger observations. For example, the electromagnetic (EM) follow-up of GW170817, has put constraint on the maximum mass of a cold, non-rotating \ac{NS}\citep{LIGOScientific:2017zic}. However, the exact nature of the remnant of GW170817 remained unknown.
%, EM follow-ups for this event could constrain the maximum mass of a cold, non-rotating \ac{NS}. 
	
	In general, the remnant of a \ac{BNS} merger depends on the mass of the merging \ac{NS} and the EoS \citep{Baumgarte_2000}. Numerical relativity simulations have indicated that the remnant can either promptly collapse to a black hole or can survive as a stable or a quasi-stable differentially rotating \ac{NS} \citep{PhysRevD.61.064001} (see \citep{Sarin:2020gxb} for review). The post-merger remnant (f-modes) oscillation mode emits gravitational waves,  which depend on its tidal deformability and compactness. The dominant frequency $f_{\text{peak}}$ is due to the excitation of the fundamental quadrupolar fluid mode $l=m=2$ \citep{fpeak1, fpeak2} and is a generic feature of all post-merger remnants \citep{Shibata}. Measuring $f_{\text{peak}}$ can potentially constrain the hot EoS models and allow us to probe the existence of exotic physics in the supranuclear density limit, e.g., temperature-dependent phase transitions. The $f_{\text{peak}}$ for a typical proto-\ac{NS} lies between $\sim$2-4 kHz, where the current and next generation of GW detectors {(i.e., LIGO-VIRGO-KAGRA (LVK) detectors)~\cite{LIGOcite, VirgoCite, KAGRA_cite}} are less sensitive. In \citep{10.1093/mnras/stad1556,  PhysRevD.100.044047}, the authors have shown that one can confidently detect these post-merger signals if the post-merger signal-to-noise ratio (SNR) is greater than eight, which is expected in the next generation of GW detectors.
	
	One of the ways to characterize the \acp{NS} from these post-merger signals relies on using hierarchical Bayesian inference (HBI) with phenomenological models that relate the chirp mass of the merging \ac{BNS} system and $f_{\text{peak}}$ of the remnant with an EoS proxy parameter (e.g., the radius of a cold \ac{NS} with mass $1.6 M_{\odot}$, $R_{1.6}$).  \citet{Criswell} used a set of simulated in-spiral-post-merger signals from \ac{BNS} coalescence to infer the injected $R_{1.6}$ with reasonable accuracy. 
	
	In this work, we extend this idea to infer radii of four different \ac{NS} masses $\{1.2, 1.4, 1.6, 1.8\}M_{\odot}$ for injected signals for different EoSs using the A+ noise curve~\cite{abbott2020noise}. We show that we can combine these four HBI inferences by assuming that in Nature, all \acp{NS} follow the same EoS.  We show that this leads to tighter bounds on the inferred \ac{NS} radii. With the piecewise-polytrope EoS model, the radii of \ac{NS} can be constrained to within $\sim 1$ km using post-merger detections in the A+ era~\cite{abbott2020noise}.
		
We organize the work as follows:
In section (\ref{sec:HBI}), we briefly describe the hierarchical Bayesian inference (HBI) proposed in \citet{Criswell} to infer the posteriors on $R_{1.6}$ and our idea of combining the posteriors of $R_{1.X}$ in HBI. In section \ref{sec:injection},  we summarize the injection events, EoSs, and priors used in the study. In section (\ref{sec:results}), we discuss the results and implications of our study. Finally, we discuss caveats and future extensions of our study in section (\ref{sec:conclusion}).

	\section{Hierarchical Bayesian Inference}
	\label{sec:HBI}

	In this section, we describe the basic idea of HBI. We review the HBI in the context of \citet{Criswell}, where the authors constrained the EoS proxy parameter, $R_{1.6}$,  using simulated events.  We provide the context of our extension of this analysis to the EoS model parameters.
	
	To understand the basic idea behind HBI, consider an astrophysical system (e.g., compact binaries) with parameters $\vec{\theta}$. Suppose we perform N independent observations on this system (labeled by $g_i$). Using Bayesian inference, we can combine these observations to give a posterior distribution on $\vec{\theta}$, as
\begin{align}
   p(\vec{\theta}|g_1\ldots g_N) = \frac{\prod_{i=1}^{N} p(g_i|\vec\theta)}{p(g_1\ldots g_N)}p(\vec{\theta}).
\end{align} 
Now, consider one observation from N different systems, then each observation will have different source parameters $\vec{\theta}_i$ ($i=1\ldots N$). If these source parameters are samples of a population,  characterized by the "population" parameter $\phi$, we can write the multi-dimensional posterior ($p(\vec{\theta}_i,\phi|d_i)$)  distribution as

\begin{align}
	p(\vec{\theta}_i,\phi|d_i) & = \frac{p(d_i|\vec{\theta}_i)p(\vec{\theta}_i|\phi)p(\phi)}{p(d_i)}. 	
\end{align}
If we combine information from N independent observations, we get the joint posterior distribution as

\begin{align}
	p(\vec{\theta}_1,\ldots \vec{\theta}_N, \phi|d_1\ldots d_N)
			& = \left(\prod_{i=1}^{N}\frac{p(d_i|\vec{\theta}_i)p(\vec{\theta}_i|\phi)}{p(d_i)}\right)p(\phi). 
\label{eq:hbi}	
\end{align}
We can marginalize over  $\theta_i$, to get
\begin{align}
p(\phi|d_1\ldots d_N) =  \left(\prod_{i=1}^{N}\vec{\int} \frac{p(d_i|\vec{\theta}_i)p(\vec{\theta}_i|\phi)}{p(d_i)}d\vec{\theta}_i\right) p(\phi)
\label{eq:hbif} 
\end{align}
This way of inferring the posterior of the "population" parameter using the information from the detected systems is termed HBI.

\subsection{Review of HBI in \citet{Criswell}}
In this section,  we describe how HBI formalism was used in \citet{Criswell} to find the expression for the posterior of $R_{1.6}$
for the astrophysical system of \ac{BNS} merger, with a post-merger remnant.
As a first step,  the authors of \cite{Criswell} used the simulation data of \ac{BNS} system from \cite{Vretinaris} which contains the peak frequency of the post-merger remnant $f_{\text{peak}}$,  and proposed an empirical relation which relates the $f_{\text{peak}}$ with the binary chirp mass  $\mathcal{M}$ and the EoS proxy parameter $R_{1.X}$ for \ac{BNS} as follows:
\begin{align}
		f_{\text{peak}} = b_0 + b_1\mathcal{M} + & b_2\mathcal{M}^2 + b_3R_{\mathrm{1.X}}\mathcal{M} \nonumber \\
		&+ b_4R_{\mathrm{1.X}}\mathcal{M}^2 + b_5R_{\mathrm{1.X}}^2\mathcal{M}. \label{EmpiricalRelation}
\end{align}
Here, the best fit coefficients, $\mathbf{b}:=(b_0,\ldots b_5)$,  are drawn from the multidimensional $\mathcal{N}(\boldsymbol{\mu}, \boldsymbol{\Sigma})$.  Further,  each \ac{BNS} observation ($d_i$) is split into inspiral ($D_{IN, i}$) and post-merger ($D_{PM, i}$) part such that the chirp mass ($\mathcal{M}$)  and the peak frequency ($f_{\text{peak}}$) are estimated independently from the inspiral and post-merger phases, respectively.

With this background,  
the authors of \cite{Criswell} mapped the HBI formulation as follows: $\phi = R_{1.X}$,  $\vec{\theta}_i = [f_{\text{peak}, i}, \mathcal{M}_i]$. Using Eq. (\ref{eq:hbif})in the HBI formalism,  the posterior on $R_{1.X}$  in terms of the likelihood contributions from the inspiral ($p(D_{IN,i}|\mathcal{M}_i) $) and post-merger ($p(D_{PM,i}|f_{\mathrm{peak},i})$) phases is
\begin{widetext}
\begin{align}
p(R_{1.X}|{\mathbf D}) = \left(\prod_{i=1}^{N} \int \int \frac{p(D_{IN,i}|\mathcal{M}_i) p(D_{PM,i}|f_{\mathrm{peak},i}) p(\mathcal{M}_i, f_{\mathrm{peak},i}|R_{1.X})}{p(d_i)} d\mathcal{M}_i d f_{\mathrm{peak},i}\right) p(R_{1.X}),
\end{align}
where $\mathbf{D}:=(d_1,\ldots d_N)$.
Further, dividing the $f_{peak, i}$ integration in discrete sum with $f_{peak, i}$ drawn from $\mathcal{N}(\boldsymbol{\mu}, \boldsymbol{\Sigma})$,  using the phenomenological relation (Eq.~(\ref{EmpiricalRelation})), and splitting the $\mathcal{M}_i$ into $I$ and $O$ regions\footnote{The domain $I$ corresponds to those $\mathcal{M}_i$ values, which yield $f_{\mathrm{peak}, ij}$ in the range [1.5, 4] kHz otherwise it belongs to $O$ domain.},  the authors get Eq. (22) of \cite{Criswell}  as
\begin{align}\label{eq:finalpost_propto}
			p(R_{\mathrm{1.X}}|{\mathbf D}) \propto p(R_{\mathrm{1.X}}) \prod_{i=1}^N \bigg[ \int_{\mathcal{I}} \frac{1}{N_{\mathrm{R}}}\sum_j\bigg(\frac{p(\mathcal{M}_i|D_{\mathrm{IN},i})}{p_0(\mathcal{M})} \frac{p(\hat{f}_{\mathrm{peak},ij}|D_{\mathrm{PM},i})}{p(\hat{f}_{\mathrm{peak},ij})}\bigg) p(\mathcal{M}_i)d\mathcal{M}_i +
			\int_{\mathcal{O}}\bigg( \frac{p(\mathcal{M}_i|D_{\mathrm{IN},i})}{p_0(\mathcal{M}_i)}\bigg) p(\mathcal{M}_i)d\mathcal{M}_i \bigg].
		\end{align}
	\end{widetext} 
	
The integration uses a grid evaluation technique, which is available as a python package {\tt BAYESTACK}~\cite{bayestack_github}. The choice of distributions for different components of Eq.~(\ref{eq:finalpost_propto}) is detailed in the appendix (\ref{sec:bast}). We call $p(R_{\mathrm{1.X}}|{\mathbf D})$ as "BaSt posterior" in this paper.
	
The authors used the catalog created in \citet{Petrov_2022}, that relied on astrophysical distributions (computed in \citet{KAGRA:2013rdx}) of mass ($\mathcal{N}(1.33 M_{\odot}, 0.09 M_{\odot})$), spin ($U(0, 0.05)$) and distance of the \ac{BNS} systems, to carry out an injection study. The authors use the results from the injections of BNS waveforms in the noise generated from the theoretical noise curve sensitivities of the advanced and A+ (projected O5)  noise curves of advanced LIGO detectors and advanced Virgo detectors~\cite{abbott2020noise}.There were three sets of injections: {\it Set A} contained simulated 26 \ac{BNS} events in one year of simulated LIGO Advance noise curve,  {\it Set B} was {\it Set A} plus the simulated 83 events in simulated A+ noise curve of one-year duration, and {\it Set C} contained expected 357 events in simulated A+ noise with four-year duration assuming \ac{BNS} merger rate density of  $320^{+490}_{-240}$ Gpc$^{-3}$yr$^{-1}$ \footnote{These events excluded the systems that would have collapsed directly to black holes (i.e., $M_{tot} > 3 M_{\odot}$, as a crude approximation)}. 
	
\citet{Criswell} created a library of GW  for limited \ac{BNS} systems using \ac{SPH} based relativistic hydrodynamical simulations \citep{SPH_1, SPH_2, SPH_3}.  It contained simulated waveforms from three different EoS curves, namely, SFHx \citep{SFHx_paper} (132 simulations), SLy4 \citep{SLy4_paper} (41 simulations), and DD2 \citep{DD2_paper} (41 simulations) with various masses. The injection waveforms for the simulated events described in {\it Sets} A,, B, C  were then obtained by nearest-neighbor approximation (in the $m_1-m_2$ space) to the library of waveforms. These approximate waveforms were then injected in respective noise and recovered using \texttt{BayesWave}, that provided the $f_{\mathrm{peak}}$ samples for $p(\hat{f}_{\mathrm{peak},ij}|D_{\mathrm{PM}, i})$ (publicly available in \citep{zenodo_criswell}).

\subsection{Our approach to recover the EoS using HBI}
\label{subsec:REoS}

Our work is built on \cite{Criswell} as follows.
In this work, we compute the $R_{1.X}$ posteriors for different radii priors using the HBI formulation prescribed in \cite{Criswell}. We further extend this formulation to obtain the estimate on the \ac{NS} EoS parameter and thus use this information to infer the tidal deformability and further constrain the M-R curves for different EoS as described below.

We assume that all \acp{NS} follow the same EoS in nature.
%We then impose an additional constraint by demanding that these four radii should be coming from the same equation of state. 
We apply this assumption through a common EoS model (for the cold \ac{NS}),  with parameters $\vec{\Upsilon}$.  Now,  we successively apply the HBI formulation by mapping the $\theta_i$s  to $R_{1.X}$ and $\phi$ to the EoS model parameters.  Using Eq. (\ref{eq:hbi}), we write

\begin{align}
	p(R_{1.2}, \ldots R_{1.8}, \vec{\Upsilon}|{\cal D})
			& = p(\vec{\Upsilon}) \prod_{i=1}^{4} \frac{p(d_i|R_{1.X_i})p(R_{1.X_i}|\vec{\Upsilon})}{p(d_i)},
\label{eq:hbi2}
\end{align}
where we consider $R_{1.X_i}$ with $X_i \in [2, 4, 6, 8]$ and ${\bf {\cal D}} \coloneqq \{d_1, d_2, d_3, d_4\}$. The likelihoods $p(d_i|R_{1.X_i})$ are the posteriors, $p(R_{1.X_i}|{\mathbf D})$, obtained from Eq.  (\ref{eq:finalpost_propto}).

We note that if $\hat{R}(\vec{\Upsilon}, 1.X_i)$ is the radius of the \ac{NS} of mass $1.X_i M_{\odot}$ computed using the Tolman-Oppenheimer-Volkoff (TOV) equations with EoS curve, defined by $\vec{\Upsilon}$, then $p(R_{1.X_i}|\vec{\Upsilon}) = \delta (R_{1.X_i}-\hat{R}(\vec{\Upsilon}, 1.X_i))$, where $\delta$ is the standard dirac delta function. Now,  following (\ref{eq:hbif}) and marginalizing the Eq. (\ref{eq:hbi2}) with respect to $R_{1.X_i}$ under HBI formulation, we get the posterior on the EoS parameter, $p(\vec{\Upsilon}|{\cal D})$ as follows:

\begin{align}
	p(\vec{\Upsilon}|{\cal D})
			& \propto \left(\prod_{i=1}^{4}\int p(d_i|R_{1.X_i})\,p(R_{1.X_i}|\vec{\Upsilon})\,d R_{1.X_i}\right) p(\vec{\Upsilon}) \nonumber \\
			& \propto \left(\prod_{i=1}^{4} p(d_i|\hat{R}(\vec{\Upsilon}, 1.X_i))\right) p(\vec{\Upsilon}),
\label{eq:hbi3}
\end{align}
where, we have used $p(R_{1.X_i}|\vec{\Upsilon}) = \delta (R_{1.X_i}-\hat{R}(\vec{\Upsilon}, 1.X_i))$.

	\begin{figure}
	 \includegraphics[trim={1cm 0 0 0},clip,scale=1]{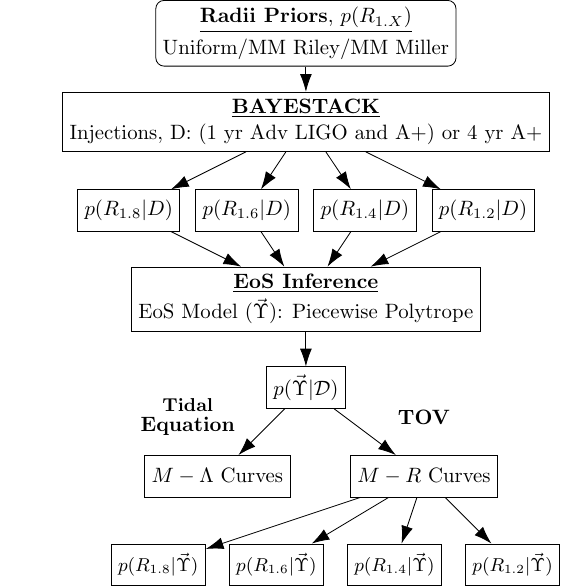}
	 \caption{Flowchart explaining the methodology.}
	 \label{fig:flow}
	\end{figure}

Figure (\ref{fig:flow}) summarizes our approach.  We start by choosing the prior, $p(R_{1.X})$, as the input to the \texttt{BAYESTACK} package~\cite{bayestack_github} that computes the HBI posteriors from Eq. (\ref{eq:finalpost_propto}),  for the four masses and for the set of simulated GW signals, $D$.  Once these posteriors are computed for the four masses,  we combine them using the Eq. (\ref{eq:hbi3}) after assuming an EoS model (characterized with the parameter $\vec{\Upsilon}$).  The posterior, $p(\vec{\Upsilon}|\mathcal{D})$, computed in this inference, is then used to compute the constraints on the mass-radius and mass-tidal deformability curves. Finally, different slices of the constrained mass-radius curves give the  $p(R_{1.X}|\vec{\Upsilon})$.
	
We draw two direct inferences: {\it First,},  successive application of the HBI formulation can help 
constrain the EoS parameter by combining inferences of the $R_{1.X}$ of \ac{NS} from different \ac{BNS} post-merger observations.  {\it Second,} mapping these constrained EoS parameters to M-R and $\Lambda-M$ curves allows us to constrain these curves. One can further take slices at $M=1.X M_{\odot}$ from the constrained M-R curves to obtain bounds for $R_{1.X}$. These bounds are found to be much tighter when compared with BaSt posteriors.

In the next section, we demonstrate the above two inferences  
using the injection study.
	
\section{Injection Study}
	\label{sec:injection}

\begin{figure}[htb!]
	 \includegraphics[scale=1]{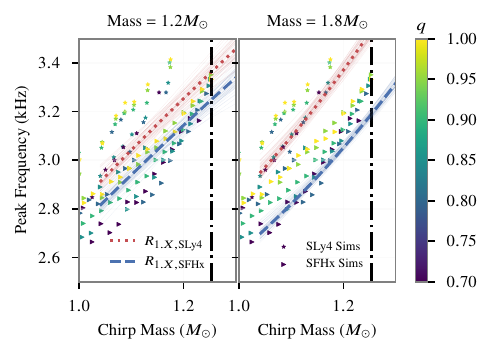}
	 \caption{Comparing the empirical relation for $f_{\mathrm{peak}}$ (denoted by blue dashed (red dotted) line for SFHx (SLy4) EoS) for the two masses with the values for SFHx (triangular marker) and SLy4 (asterisk marker) obtained using currently available waveforms. The thin lines near the best fit empirical curves denote 100 samples obtained from the covariance matrix of the fit. The vertical dot-dashed line corresponds to the maximum chirp mass from the set of available waveforms for SFHx EoS-based systems.} 
	 \label{fig:Emp_fit}
	\end{figure}

In order to understand how well the empirical relation fits the injections studied in the \citet{Criswell}'s study,  we plot peak frequency {\it vs} chirp mass in Figure (\ref{fig:Emp_fit}) for SPH based simulated waveforms, along with the empirical curves. 
 
 The figure uses  simulated (using \ac{SPH}) \ac{BNS} system having SFHx (132 simulations denoted by triangles) and SLy (41 simulations denoted by asterix) EoS curves.  It further plots $f_{\mathrm{peak}}$ {\it vs} chirp mass for the systems with $R_{1.2}$ and $R_{1.8}$ for the best-fit parameters (dotted red line for SLy4 and dashed blue line for SFHx) from the empirical relation given by Eq. (\ref{EmpiricalRelation}). The samples for the fit are drawn using the covariance matrix (thin blue for SFHx and red lines for SLy4) of the empirical fit. We compare these $f_{\mathrm{peak}}$ curves with the values obtained from the \ac{SPH} simulations for SFHx (triangles) and SLy4 (asterisk) EoS.  We find that the values $f_{\mathrm{peak}}$ values agree better with the empirical relation for the $R_{1.2}$ case compared to $R_{1.8}$ case.
 
 We draw a black vertical dash-dotted line at 1.254 $M_{\odot}$, the upper bound on the chirp mass from the \ac{SPH} simulated systems for SFHx EoS.  We find that 157 injected events  in {\it Set C} have chirp mass $> 1.254 M_{\odot}$ (beyond the black dash-dotted line in the plot),  and thus making the SPH simulated waveforms insufficient to capture these injected systems. Thus,  extrapolation procedure during waveform computation,  that uses the nearest neighbour approach
 described in Sec.\ref{sec:HBI}.  
  makes it unreliable.  This is a potential source of bias in our analysis. 
  
In summary,  we choose the following subset of injection sets to carry out our HBI study and to perform the bias study using the waveform interpolations.  The first two sets are {\it Set B} with SFHx and SLy EoS. {\it Set C} with SFHx EoS is the third set. The fourth set, {\it Set D}, contains 200 events from {\it Set C} with chirp mass less than 1.254  $M_{\odot}$.
We prepare this set especially to understand and compare the interpolation bias from our results. We use the available $f_{\text{peak}}$ posteriors library \citep{zenodo_criswell} for this purpose. 

\begin{figure*}[htb!]
		 \includegraphics[scale=1]{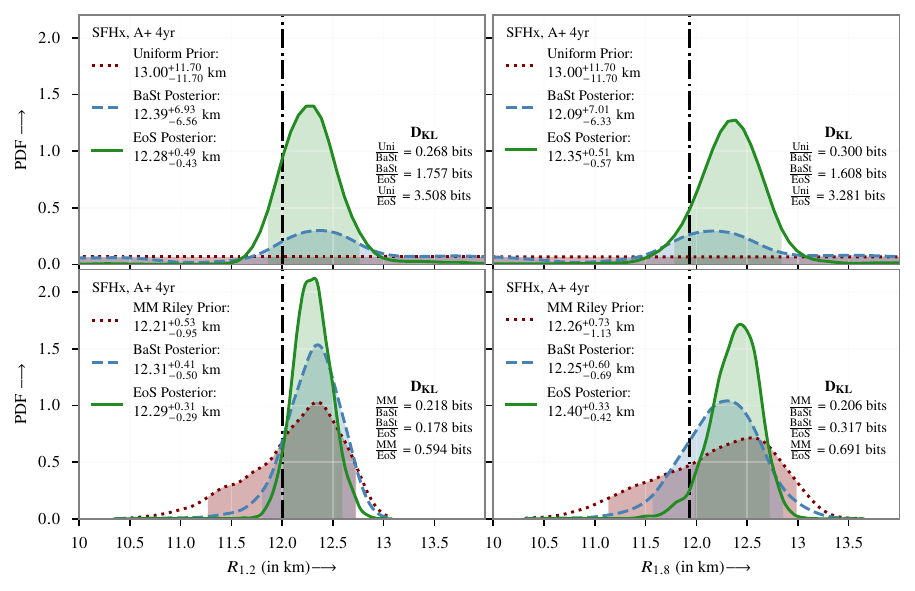}
		 \caption{Priors and posteriors are plotted for the radii of cold \ac{NS} with masses $1.2 M_{\odot}$ (left) and $1.8 M_{\odot}$ (right). The red dotted curves in the bottom row correspond to analysis performed with astrophysical MM Riley prior on radii. The top row plots the distribution for the same quantities but now starting with a uniform prior on radii. The blue dashed curves represent posteriors, $p(R_{1.X}|D)$, computed using \texttt{BAYESTACK} (BaSt) with simulated events in {\it Set C}. Green solid curves represent posterior computed from Bayesian inference after combining the four $p(R_{1.X}|D)$s using the piecewise polytropic model. Here, the injected EoS is SFHx with dot-dashed vertical lines representing the radii of \ac{NS} for the respective masses. Also computed are the Kullback -- Leibler divergences ($D_{KL}$) between the different distributions and the median and 90\% credible intervals for the radius distributions.} 
		 \label{fig:Aplus}
	\end{figure*}

\subsection{EoS Model}
	\label{sec:EoSM}

In this subsection, we discuss the EoS model that we use in this study.  We use the piecewise polytrope EoS model \cite{Read_2009} with a fixed crust. The crust is modeled by four polytropic pieces, whereas we use three polytropes (with varying parameters) for the inner core. Each polytropic piece is given by, 
	\begin{equation}
		P(\rho) = K_i \rho^{\Gamma_i},
		\label{polytrope:definition}
	\end{equation}
	where $P$ is pressure,  $\rho$ is the rest mass density, and $K_i$,  $\Gamma_i$ are model parameters. We use the analytic form of SLy4 \citep{SLy4_paper} (Table II of \citet{Read_2009}) to model the crust. For the core, we parameterize the EoS curves using $\Gamma_5$, $\Gamma_6$, $\Gamma_7$, and $\log{P_5}$ (i.e. $\vec{\Upsilon} = [\log{P_5}, \Gamma_5, \Gamma_6, \Gamma_7]$). We compute the $K_i$ in the expression above by demanding the continuity of the pressure at the transition density points, with densities given by $10^{14.7}  \,\mathrm{g/cm^3}$ and $10^{15}\,\mathrm{g/cm^3}$.   
	
\subsection{Priors}
	\label{sec:inputs}
We explore the effects of radius prior, $p(R_{1.X})$ (for Eq. (\ref{eq:finalpost_propto})) on the BaSt and EoS posteriors. The first one is the uniform, $U(9 \text{km}, 15 \text{km})$. The second and third are multimessenger priors that are obtained from astrophysical observations, as follows. We use the mass-radius posteriors obtained using the GW1710817, and calculations~\cite{Miller_1,Miller_2, Riley_1, Riley_2} based on the NICER light curves from PSR J0030+0451 and PSR J0740+6620 \citep{Fonseca_2021} (see \citet{Tiwari_2024} for details). Using the analysis done by \citet{Tiwari_2024} on the calculations by Miller et al. ~\cite{Miller_1,Miller_2} and Riley et al. ~\cite{Riley_1, Riley_2}, we get two mass-radius posteriors. The distribution corresponding to Miller et al. ~\cite{Miller_1,Miller_2} calculations is stiffer than the one obtained via Riley et al. ~\cite{Riley_1, Riley_2} calculation. We refer to them as multimessenger (MM) Miller/Riley. Please note that the chosen astrophysical priors are different from those of \cite{Criswell}. 

The priors on the EoS parameters, $\vec{\Upsilon}$, are $\Gamma_4 \sim U(1.4, 5.0)$, $\Gamma_5 \sim U(1.0, 5.0)$, $\Gamma_6 \sim U(1.0, 5.0)$ and $\log{P_5} \sim (33.5, 35.5)$.  
% For the spectral model, we refer to the \emph{high-$p_0$ spectral model} in \citet{Carney_2018} and use $p_0 = 5.4\times10^{32}\, \mathrm{dyne \, cm^{-2}}$ with priors \citep{Abbott_2018}: $\gamma_0 \in  [0.2, 2]$, $\gamma_1 \in  [-1.6, 1.7]$, $\gamma_2 \in  [-0.6, 0.6]$, and $\gamma_3 \in [-0.02, 0.02]$.
We ensure that the EoS curve for each $\vec{\Upsilon}$ can support $1.8 M_{\odot}$ \ac{NS}. Additionally, we fit the beta equilibrated EoS curves of SFHx and SLy4 to the above EoS model to ensure that the best-fit parameter values for the injected EoS curve are inside the prior range for $\vec{\Upsilon}$ (see appendix (\ref{appx:fitting1}) for the fit values).

	\section{Results}
	\label{sec:results}
	We summarize the main results from the injection study.
		\begin{figure*}[htb!]
		\includegraphics[scale=1]{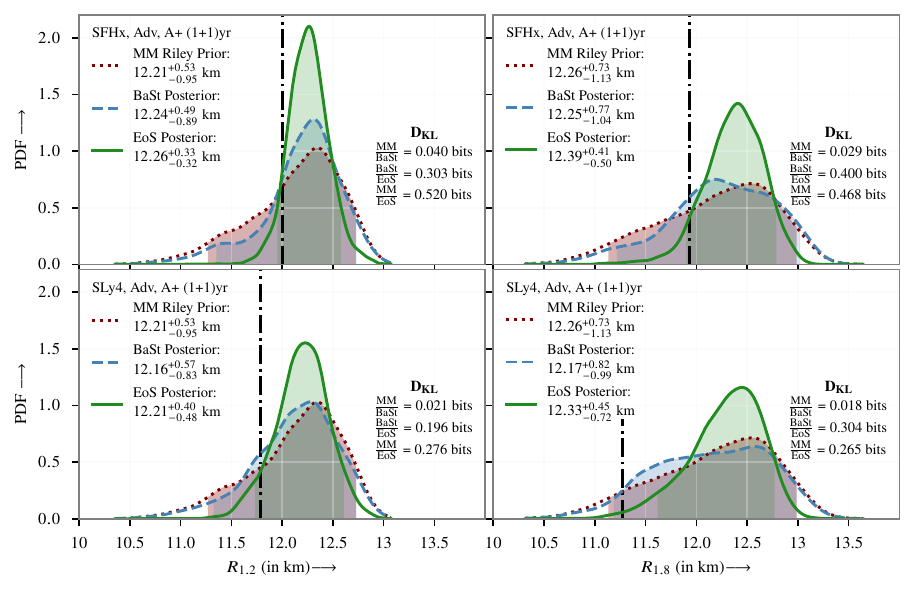}
		\caption{Similar to Figure (\ref{fig:Aplus}). Here, we show the radii distribution corresponding to events in {\it Set B}. Injected EoS for the top row is SFHx, while for the bottom row, it is SLy4.} 
		\label{fig:O4O5}
	\end{figure*}
	\subsection{Effect of Combining BaSt Posteriors}
	
	In Figure (\ref{fig:Aplus}), we present the constraints on the radius of cold \acp{NS} with masses $1.2 M_{\odot}$ and $1.8 M_{\odot}$.   In the top (bottom) panels of the  Figure (\ref{fig:Aplus}), we use a uniform (astrophysically informed MM Riley) prior on the radii, as denoted by the red dotted curves.  The first set of posteriors (denoted by dashed blue curves) are BaSt posteriors using injection {\it Set C}. The second set of radii posteriors (denoted by solid green curves) are obtained from performing HBI on the four BaSt posteriors for the common piecewise polytrope EoS model.
		
When we compare the bottom panels of posteriors of  $R_{1.8}$ and $R_{1.2}$,  we find that the informed prior on radii reduces the $90\%$ credible interval (CI) of the $R_{1.X}$ radii distribution by $\sim 34\%$,  if we rely only on the {\texttt{Bayestack}} technique.  Additionally,  when we impose the common EoS model across the four radii,  there is a further reduction in the $90\%$ CI of $R_{1.X}$ by $\sim 38\%$, thus recovering the injection within $\sim 0.6$ km. 

We further find that in all the $R_{1.X}$ cases,  the posterior distributions shift from the peak of the BaSt prior (red dotted lines) towards the injected $R_{1.X}$ value.  This clearly shows the additional constraint due to common EoS and thus provides additional information to bring it closer to the injection value.  But as the mass increases from $1.2 M_{\odot}$ to $1.8 M_{\odot}$, the shift towards the injection reduces. In Figure (\ref{fig:Aplus}), we present the two cases to demonstrate the best and worst scenario. In the best  $R_{1.2}$ case,  the shift towards the injected value is significant as compared to $R_{1.8}$ case.   The latter might point to certain systematics in the analysis, which we discuss later in the text.

The top panel of the Figure (\ref{fig:Aplus}) shows that the radii of the injected EoS can be recovered within 1 km ($90\%$ CI) when no astrophysical prior is imposed. 

 % these two figures are now placed side by side for correct float placement by Latex

%\begin{figure}
% \includegraphics[scale=1]{figs/Waveform_Interpolation_Accuracy}
% \caption{Plotting the .} 
% \label{fig:Wave_Int}
%\end{figure}
We repeat the same analysis with astrophysical MM Riley prior and simulations with {\it Set B} (i.e., 109 events). 
In the top panel of Figure (\ref{fig:O4O5}),  we present the results. In this case, there is
$\sim  7\%$ reduction in the
 $90\%$ credible interval (CI) of the $R_{1.X}$ radii distribution compared to the MM Riley prior.  However, imposing a common EoS constraint, we get an overall improvement in CI by $\sim$ 50-60\%.  Our analysis with all $R_{1.X}$ shows that we recover the injected radii within 90\% CI for all cases except for $R_{1.8}$.  This deviation is possibly
due to less support for the injected value in the prior distribution and the bias introduced due to the validity of the empirical fits in the region of simulated injections. 

The bias in the recovery of the injected radii in Figure (\ref{fig:Aplus}) and ({\ref{fig:O4O5}) might be pointing to certain systematics in the analysis due to the intermediate BaST posterior, namely the validity of the empirical fits in the region of simulated injections
and the waveform interpolation error. 
The bias in the recovery increases with increasing mass of the radius slice.   We find that the empirical relation becomes less reliable in computing the 
accurate  $f_{\mathrm{peak}}$ value for higher chirp mass and mass ratio. This situation worsens for $R_{1.8}$ for the SFHx case.  Since we use the empirical relation in obtaining the BaSt posterior,  this inaccuracy manifests as bias in the recovery.  The bias component due to waveform interpolation is due to 
 associating the injected waveform with the closest available waveform from \ac{SPH} simulations in $m_1-m_2$ space.  We find minimal overlap between the available waveforms and injected events for the high chirp mass systems as discussed in Sec.\ref{sec:injection}. This implies that the chosen waveforms for the high chirp mass events carry the features of low-mass ones, thus compromising the injection analysis.  However, these are systematics of the injection study and do not affect inference from real data (as opposed to the bias from the empirical relation, which needs improvement in the future).  This means that our inference, in this case,  will be broadly affected by the astrophysical priors and the EoS model.  
 
 %In figure (\ref{fig:Wave_Int}), we plot the available waveform from SPH along with the sampled events. 
	
	We have also computed the K -- L divergences between different radii distributions to see if we are improving our radii estimates. We find that the information gained from BaSt inference monotonically increases with increasing \ac{NS} mass. In contrast, for the EoS inference, there is a sharp drop in the information gain when going from $R_{1.6}$ to $R_{1.8}$ for the uniform prior. This can be attributed to the inaccuracies in the empirical relation at high mass. 	%\hl{(mention values? like how sharp it is)}

	\subsection{Inference with different injected EoS}

	We vary the injected EoS to investigate its plausibility for its recovery in the HBI framework.  We use SLy4  and SFHx EoS in this study with injections from {\it Set B} (109 simulated events). We choose {\it Set B} to be commensurate in terms of the available \ac{SPH} waveforms and the corresponding simulated events. We present the results in Figure  (\ref{fig:O4O5}) with top(bottom) panels corresponding to SFHx (SLY4) EoS using astrophysical MM Riley prior.

We find that $90$ \% of the CI of the radii distribution corresponding to SFHx EoS are $\sim 27$\% smaller than that of the SLy4 EoS. This can be attributed to two factors, namely, prior support to the injected value and accuracy of the empirical relation for the two EoSs.  For the SLy4 EoS injections,  the radii prior have less support for the injected value and thus shows poor recovery in radii.  The Figure (\ref{fig:Emp_fit}) shows that the empirical relation approximates SFHx EoS better than SLy4 EoS.  This makes the final posterior for $R_{1.2, \mathrm{SLy4}}$ broader than $R_{1.2, \mathrm{SFHx}}$. 

	\subsection{Constraining Tidal Deformability and Radius}
	\begin{figure}
		 \includegraphics[scale=1]{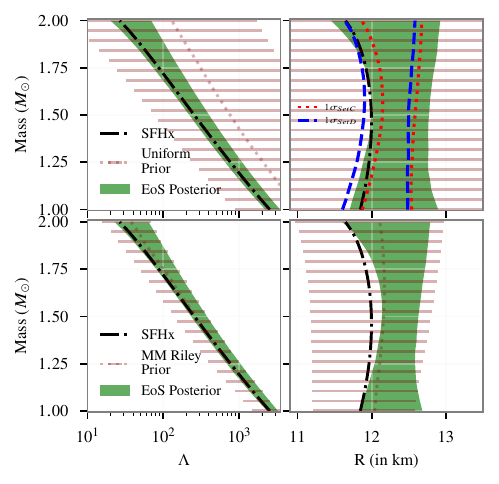}
		 \caption{Constraints on the M -- $\Lambda$ (left) and M -- R (right) curves from the simulated post-merger events from {\it Set C}. The inference in the top (bottom) row corresponds to uniform (MM Riley) BaSt priors for $R_{1.X}$. Here, red error bars represent the 90\% CI range for the priors on $R_{1.X}$; green patches represent the 90\% CI bounds obtained from the $\vec{\Upsilon}$ posteriors, and black dot-dashed lines represent the curves obtained from the injected EoS, i.e., SFHx. The top right subfigure additionally plots, with red dotted (blue dashed), $M-R$ 1-$\sigma$ bounds using the Set C (Set D) events.} 
		 \label{fig:MRLam}
	\end{figure} 
	
Here,  we investigate the recovery of the injected EoS in the M - $\Lambda$ and M-R space using {\it Set C} and {\it Set D}. In Figure \ref{fig:MRLam}, we plot the uncertainties in M-$\Lambda$ (left column) and M-R (right column) curves obtained after constraining the EoS parameters using both the priors during the BayesStack analysis. The results with the uniform (astrophysically informed MM Riley) radii prior are in the top (bottom)  rows.  We plot  $1\sigma$ CI on M-R curves (top-right panel) from signals in {\it Set C} (red-dotted) and {\it Set D} (blue-dashed)

We find that {\it Set C} can recover the injected M - $\Lambda$ curve at the 90\% CI boundary for both the priors. However,  this is not the case for the recovery of injected EoS in terms of the injected $M-R$ curve. While using both MM Riley and MM Miller priors, the injected EoS falls outside the 90\% bound in $M-R$ space.  In Figure.   \ref{fig:MRLam}, we show the results with MM Riley prior; however, MM Miller shows similar results. We find for both astrophysical priors that, as the injected radius goes away from the peak of the prior distribution, the final posterior is less likely to capture the injected value within a 90\% CI.  

For uniform prior, we capture the injected $M-R$ curve within 90\% CI, although there is an inherent bias in the recovery.   In this case,  the 90\% CI of the \ac{NS} radius is $\sim$ 1 km.   The {\it Set C} of post-merger events in A+ may constrain the radii of \ac{NS} (all the four masses corresponding to $R_{1.X}$} considered in this work) to within 1 km.  To investigate the bias incorporated by waveform interpolation,  we use a tailor-made set {\it Set D} by dropping the events with high chirp mass.  Now,  as expected,  the recovery M-R curve with {\it Set D} is better, consistent with the injected curve as compared to {\it Set C},  which
has issues with the waveform interpolation, clearly evident from the top-right panel of Figure (\ref{fig:MRLam}).
 This is because we have now removed those masses for which the empirical relation was inaccurate, and thus, we have reduced the systematics partially responsible for the bias.

%\subsection{EoS Model Bias}
	\section{Discussion and Conclusions}
	\label{sec:conclusion}
	
		In this work, we have explored the implications of observing post-merger signals from the next generation of GW detectors. The first part follows the work of ~\cite{Criswell}, where the authors developed HBI to translate the information from the GW detections with post-merger remnant to the posterior of $R_{1.6}$ through an empirical relation that expresses $f_{peak}$ in terms of the binary parameter ${\cal M}$ and the radius of the \ac{NS} $R_{1.6}$.  We used this framework to compute posteriors for $R_{1.X} (X=2, 4, 6, 8)$ for similar sets of post-merger remnant injections and astrophysical priors for $R_{1.X}$.  In the second part of this inference, we propose a new framework to combine the $R_{1.X}$ posteriors to obtain the constraints on the parameters of the piecewise-polytrope EoS model. 
		
Using an injection study with this framework,  we find that a merger rate density of $320^{+490}_{-240}$ Gpc$^{-3}$yr$^{-1}$ in the A+ era for four years of operation (yielding ~357 events) would reduce the 90\% CI bounds on radius (for cold neutron stars) uncertainty to within a km. On the other hand, if astrophysically informed priors (i.e. radii constraints coming from GW170817~\citep{GW170817_Obs} and the two NICER calculations~\cite{Miller_1, Miller_2, Riley_1, Riley_2} ) are used for radii, the radii constraints become tighter (i.e., 90\% CI bounds on radii becomes $\sim 0.55$ km).  However, we find that the recovery of the injected SLy4 and SFHx EoS across different $R_{1.X}$ priors is biased due to certain systematics.  We also study the overall impact of this bias in inferring the EoS of \ac{NS}.  We find that we can recover the injected $M-\Lambda$ curve at the boundary of the 90\% bounds, but there is a bias in recovering the injected $M-R$ curve. We infer that this bias can be attributed to the interpolation error during waveform estimation for the injected events.  Additionally, the validity of the empirical relations over the range of radii incorporates an additional source of uncertainty. Our investigation shows that the invalidity of the empirical relation in certain regions can lead to biases.  We have done preliminary bias studies with limited data, and we leave it for future work to explore and mitigate the effects of these uncertainties on the final inference. With improved empirical relations, the framework developed here would help characterize BNS systems, especially during the hot merger phase of the evolution, thereby probing the inner core of neutron stars. This becomes important in the third generation of GW detectors, where, despite limited improvement in sensitivity at the high-frequency regime, the sheer number of expected detections of BNS systems would allow us to probe their remnants using this framework. We note that any future injection study should use the updated merger density consistent with the observed BNS systems.

Using an astrophysically informed prior, we study how these biases impact the recovery across two EoS curves, SFHx and SLy4, for 109 simulated events injected in advanced LIGO and A+ noise ({\it Set B}). This has direct implications for probing the phase transition during the merger phase.   Now,  being closer to the peak of MM Riley prior,  if we treat SFHx as the cold beta equilibriated EoS and SLy4 as the hot post-merger EoS, then we should be able to use the difference in the recovery posterior in the two cases to infer the presence of a phase transition. We find that, with 109 events and the current accuracy of the empirical relations for the four radii, the comparison of the recovery posteriors remains inconclusive.  However,  with the third generation of ground-based detectors such as Einstein Telescope~\cite{Punturo_2010} and Cosmic Explorer~\cite{2019BAAS...51g..35R},  and thus with more events, the prospects of probing the phase transition can improve significantly, and we should be able to probe the EoS parameters with higher accuracy.
		
	\section*{Acknowledgements}
	We thank Debarati Chatterjee for her valuable comments. AP acknowledges the support from SPARC MoE grant SPARC/2019-2020/P2926/SL, Government of India.
The authors greatly acknowledge the use of the Utsu cluster and Infinity server hosted in the physics department of IIT Bombay, and supported under the SERB Power grant.
\\
	\begin{appendix}
	        \section{Probability distributions used for BaSt}
	        \label{sec:bast}
	        In this appendix,  we summarize the choice of certain distributions by \citet{Criswell} to get the BaSt posteriors.  In \cite{Criswell}, the authors use \texttt{BayesWave}  \citep{BayesWave_1, BayesWave_2} to obtain the posterior, $p(\hat{f}_{peak, ij}| D_{PM, i})$, from simulated GW signal. The \texttt{BayesWave} constructs the GW template using Morlet-Gabor wavelets, and the associated parameters are sampled using the detected GW signal. These parameter samples are then mapped to the peak frequency posterior.  

The choices for chirp mass-related distributions are as follows:
The priors $p_0 (\mathcal{M}) \sim U(0.4 M_{\odot}, 4.4 M_{\odot})$ and  $p(\mathcal{M})\sim \mathcal{N}(1.33 M_{\odot}, 0.09 M_{\odot})$, while the posterior
$p(\mathcal{M}_i|D_{\mathrm{IN},i}) \sim\mathcal{N}(\mathcal{M}_{i}, \sigma_{\mathcal{M}_i})$. The 
$\sigma_{\mathcal{M}_i}$ is%citation added for bayeswave
	\begin{equation}
		\label{eqn:chirp-mass-scaling}
		\sigma_{\mathcal{M}_i} = \frac{\mathcal{M}_i}{\mathcal{M}_{\textrm{170817}}}\frac{\textrm{SNR}_{170817}}{\textrm{SNR}_i}\sigma_{\textrm{170817}}
        \end{equation}
with  $\textrm{SNR}_{170817}=32.4$ is the SNR of GW170817 \citep{GW170817_Obs}.
	
	The prior for $f_{\text{peak}}$, $p({f}_{\mathrm{peak}})$, was obtained using \texttt{BayesWave} technique with wavelet parameters being sampled  following BayesWave priors \citep{BayesWave_1, BayesWave_2}. Two different priors on \ac{NS} radii were; $p(R_{\mathrm{1.X}})$ following a uniform distribution, and astrophysically constrained~\cite{Huth_2022}.
	
		\section{Fitting Parameters for EoS Curves}
		\label{appx:fitting1}
		
		We use the data available on CompOSE \citep{refId0, Compose_data}: SFHx \citep{SFHx_data}
		%DD2 \citep{DD2_data} 
		and SLy4 \citep{SLy4_data}. As we deal with cold \ac{NS} EoS, we select the energy $(\epsilon_i)$ -- pressure $(P_i)$ data for SFHx at 0.1 MeV temperature. We find the parameters using Levenberg -- Marquardt minimizing algorithm on the residual (defined in \citet{Lindblom_2010}),
		\begin{equation}
			\Delta^2(\vec{\theta})=\sum\limits^N_{i=1}\frac{1}{N}\left[\log{\left\{\frac{\epsilon_{\mathrm{fit}}(P_i, \vec{\theta})}{\epsilon_i}\right\}}\right]^2
			\label{appx:residual}
		\end{equation}
		where $\vec{\theta}$ are the parameters of the model and $N$ is the number of data points.  We list the fitted values in the table below.
		\begin{table}[H]
			\centering
			\begin{tabularx}{\linewidth}{|Y|Y|Y|Y|Y|}
				\hline
				EOS & $\log{P_5}$ & $\Gamma_4$ & $\Gamma_5$ & $\Gamma_6$ \\
				\hline
				SLy4 & 34.3817 & 2.9823 & 2.9993 & 2.8454 \\
				\hline
				SFHx & 34.4769 & 4.2129 & 3.0264 & 2.4029 \\
				%\hline
				%DD2 & 34.6349 & 3.3491 & 3.1327 & 2.2955 \\
				\hline
			\end{tabularx}
			\caption{Fitted parameters for the piecewise polytrope model.}
			\label{appx_polytrope_table}
		\end{table}
		The fitting parameters for SLy4 have values similar to those calculated in \citet{Read_2009}.	
	\end{appendix}

	\bibliography{references.bib}
\end{document}